\documentclass[twocolumn,english,aps,pre,twocolumn,superscriptaddress,showkeys,showpacs]{revtex4-1}
\usepackage{caption}
\usepackage[T1]{fontenc}
\usepackage[utf8]{inputenc}
\usepackage{color}
\usepackage{babel}
\usepackage{array}
\usepackage{booktabs}
\usepackage{braket}
\usepackage{multirow}
\usepackage{amsmath}
\usepackage{amssymb}
\usepackage{graphicx}
\usepackage{float}
\usepackage{amssymb}
\usepackage{physics}
\usepackage{subcaption}
\PassOptionsToPackage{normalem}{ulem}
\usepackage{ulem}
\usepackage[unicode=true,pdfusetitle,
 bookmarks=true,bookmarksnumbered=false,bookmarksopen=false,
 breaklinks=false,pdfborder={0 0 1},backref=false,colorlinks=false]
 {hyperref}

\usepackage{graphicx}
\graphicspath{{figures/}}

\makeatletter

\hypersetup{
    colorlinks=true,
    linkcolor=red,
    citecolor=blue,
    filecolor=blue,      
    urlcolor=blue,
}
\makeatother

\begin{document}

\keywords{electromagnetically induced transparency, slow light, group velocity, collective EIT}

\author{H.~Sanchez}
\author{L.~F.~Alves da Silva}
\address{Instituto de Física de São Carlos, Universidade de São Paulo, Caixa Postal 369, 13560-970, São Carlos - SP, Brazil}

\author{M.~A.~Ponte}
\address{Instituto de Ciências e Engenharia, Universidade Estadual Paulista - Câmpus de Itapeva, 18409-010, Itapeva - SP, Brazil}

\author{M.~H.~Y.~Moussa}
\address{Instituto de Física de São Carlos, Universidade de São Paulo, Caixa Postal 369, 13560-970, São Carlos - SP, Brazil}

\author{Norton~G.~de~Almeida}
\address{Instituto de Física de São Carlos, Universidade de São Paulo, Caixa Postal 369, 13560-970, São Carlos - SP, Brazil}
\address{Instituto de Física, Universidade Federal de Goiás, 74.001-970, Goiânia - GO, Brazil}

\title{From superradiance to collective EIT in three-level ensembles}

\pacs{42.50.Nn, 05.70.Ln, 03.65.Yz }

\begin{abstract}

We investigate the collective dynamics of a three-level ensemble under the Dicke limit, 
revealing a unified connection between superradiant emission and electromagnetically induced transparency (EIT). 
Our results show that the transient superradiant burst exhibits the expected peak intensity scaling $I_{\max}\!\sim\! N^2$, 
with a universal finite-size correction $|\xi(N)-2|\!\sim\! 1/\ln N$ that governs the apparent scaling exponent in realistic ensembles. 
In the stationary regime, collective broadening modifies the EIT response: although it typically enhances absorption, 
it counterintuitively increases the group velocity, leading to a relative scaling $v_g\!\propto\! N^2$, 
even while $v_g\!\ll\! c$. 
This effect suggests that cooperative interactions fundamentally limit the achievable slow-light delay in dense media. To achieve these results, we derive a representative-atom master equation that quantitatively reproduces both the superradiant and EIT regimes, in excellent agreement with the exact symmetric-subspace dynamics and correctly incorporating collective feedback and $N$-dependent broadening.
This unified framework bridges transient superradiant emission and steady-state quantum interference, with direct implications for slow light, quantum memories, and precision metrology.

\end{abstract}

\maketitle

\section{Introduction} Collective light–matter interactions have long been recognized as a cornerstone of quantum optics. 
A paradigmatic example is Dicke superradiance (SR) \cite{dicke1954coherence}, where an initially inverted ensemble of $N$ emitters radiates cooperatively, producing an intense burst with a peak intensity scaling as $I_{\max}\!\sim\!N^2$ and a characteristic temporal profile. 
This transient cooperative emission has been extensively studied in two-level systems, and more recently extended to multilevel structures where additional radiative channels and interference effects emerge \cite{ito1976quasi,bowden1978cooperative, crubellier1978level,crubellier1985superradiance,bashkirov2000superradiance,malyshev2000super,johnsson2002quantum,  jin2003superradiance,bashkirov2006quantum,konovalov2020master}. 

A complementary phenomenon is electromagnetically induced transparency (EIT), where destructive quantum interference in a three-level $\Lambda$ system suppresses probe absorption at two-photon resonance \cite{fleischhauer2005electromagnetically,scully1997quantum,jin2003superradiance}. 
EIT is inherently a stationary effect, requiring a weak probe, a strong control field, and near two-photon resonance. 
It has found applications ranging from slow light and quantum memories \cite{hsiao2020mean,rastogi2022superradiance} to precision spectroscopy and quantum metrology \cite{feng2017electromagnetically}. 
In particular, in EIT-based slow-light experiments it is well established that the achievable group delay is fundamentally limited by the available optical depth and by additional homogeneous broadening mechanisms (such as buffer-gas collisions and interaction-induced dephasing), which broaden the transparency window and increase residual absorption, thereby reducing the delay~\cite{fleischhauer2005electromagnetically,hau1999light}. 
The possibility of harnessing collective superradiant emission for metrology has been experimentally explored in the context of active optical clocks \cite{bohnet2012steady}.

In this work, we analyze collective superradiance and EIT in a $\Lambda$ configuration for a moderately dense ensemble of three-level atoms. To this end, we employ a mean-field approach \cite{ito1976quasi,bowden1978cooperative,mizrahi1993pulsed,malyshev2000super,hsiao2020mean} , which provides a unified framework for treating both phenomena. In optically dense, collectively coupled $\Lambda$ ensembles, the same mean-field physics that drives superradiant emission also reshapes the steady-state EIT response, yielding quantitative performance bounds for slow light. We show that, in the Dicke limit, cooperative radiative broadening scales with $N$ and enters directly the probe coherence that defines the EIT window and dispersion. Consequently, increasing the effective optical depth by enlarging $N$ simultaneously induces cooperative broadening that narrows the EIT window and limits the achievable slow-light delay. This question differs from schemes that engineer EIT-like spectra by coupling collective super- and subradiant states in effectively two-level settings (e.g., Ref.~[16]). In particular, we isolate a purely radiative Dicke-limit mechanism---collective broadening in an otherwise noninteracting, homogeneous ensemble---rather than interaction-driven effects in multilevel ensembles \cite{manassah1998superradiance,sutherland2017superradiance,pineiro2022emergent,hao2021observation,suarez2022superradiance} or multiple scattering in extended clouds.

This framework captures both the transient SR burst and the steady-state EIT response within a single mean-field description, thereby linking cooperative emission to stationary quantum interference. 
It predicts a universal finite-size bias in the apparent SR scaling exponent, $|\xi(N)-2|\sim 1/\ln N$, and shows that cooperative radiative broadening reshapes the EIT spectrum by narrowing the transparency window and modifying its contrast. 
These results clarify how collective broadening constrains slow-light performance while providing an ultra-narrow spectral feature relevant to filtering and precision spectroscopy.

\begin{figure}[t]
\centering
\includegraphics[width=\columnwidth]{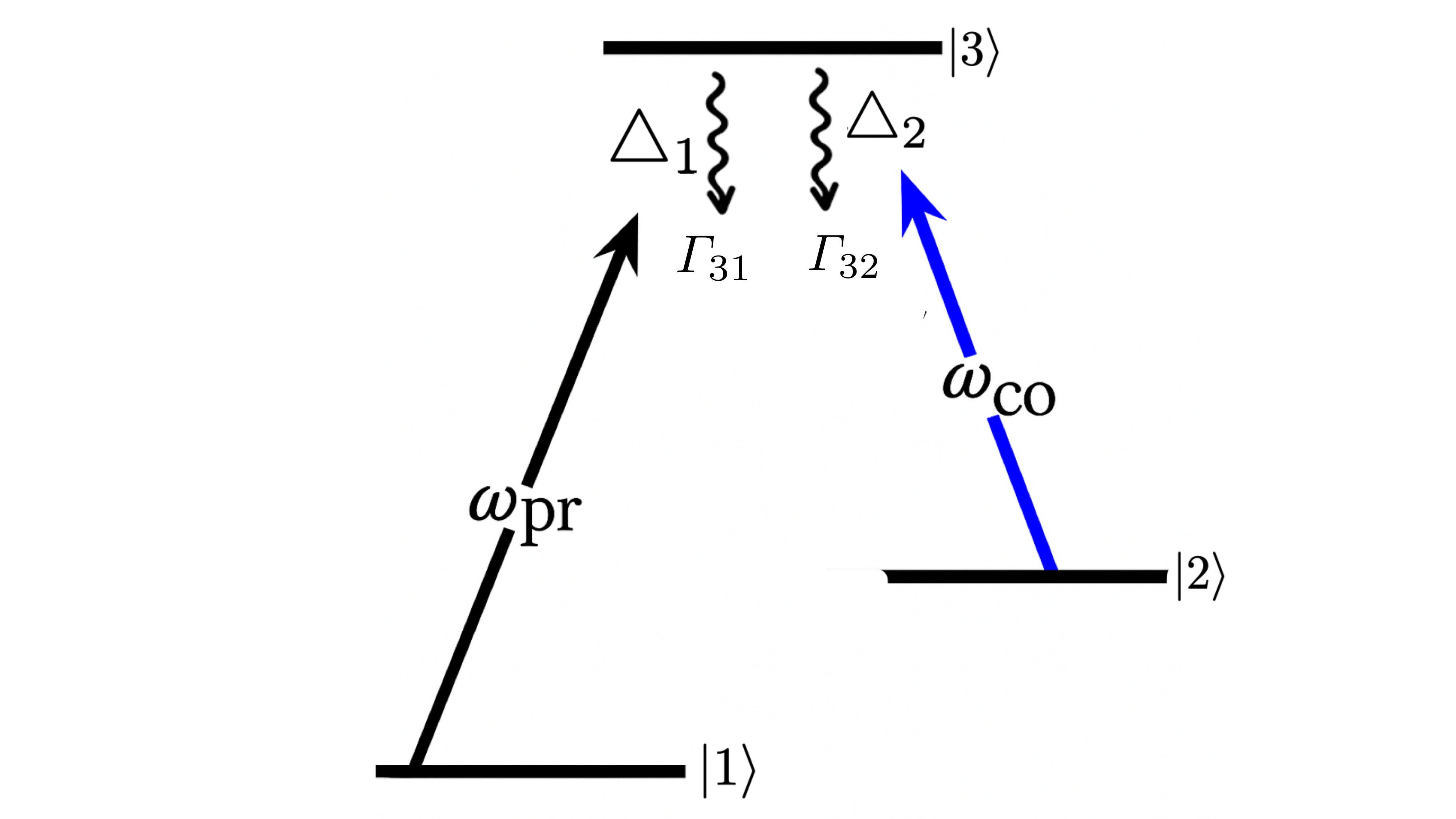}
\caption{
Three-level $\Lambda$ configuration relevant to electromagnetically induced transparency (EIT) and superradiance (SR). 
A weak probe field drives the $|1\rangle \leftrightarrow |3\rangle$ transition with frequency $\omega_{p}$ 
and detuning $\Delta_{1} = \omega_{31} - \omega_{p}$, 
while a strong control field couples $|2\rangle \leftrightarrow |3\rangle$ with frequency $\omega_{c}$ 
and detuning $\Delta_{2} = \omega_{32} - \omega_{c}$. 
$\Gamma_{31}$ and $\Gamma_{32}$ denote the spontaneous emission rates for the $3\!\to\!1$ and $3\!\to\!2$ channels.
Superradiance occurs when both fields are turned off, 
whereas collective EIT emerges in the steady state when both fields are applied.}
\label{fig:Lambda_scheme}
\end{figure}

\section{The model} We consider an ensemble of $N$ identical three-level emitters in the $\Lambda$ configuration of Fig.~\ref{fig:Lambda_scheme}, 
with long-lived states $\ket{1}$ and $\ket{2}$ coupled to the excited state $\ket{3}$ by optical dipoles $\boldsymbol{\mu}_{31}$ and $\boldsymbol{\mu}_{32}$. 
We focus on the stationary, homogeneous Dicke-limit regime ($L\!\ll\!\lambda_0$), which is directly relevant for solid-state spin ensembles, superconducting qubits, and quantum dots embedded in low-$Q$ cavities, as well as for circuit-QED platforms and sufficiently cold atomic samples, where motion-induced inhomogeneities are negligible. In this setting, the dominant additional linewidth entering the probe coherence is a \emph{homogeneous} radiative contribution associated with cooperative (collective) broadening, which directly competes with the EIT transparency width and must be distinguished from the collisional broadening mechanisms commonly discussed in optically dense slow-light experiments. Including Doppler and other inhomogeneous broadening mechanisms (e.g., warm vapor or moving cold-atom ensembles) would require an additional velocity average over the microscopic susceptibility and lies beyond the scope of the present study. We nevertheless expect the cooperative radiative mechanism identified here to remain relevant whenever collective broadening competes with the EIT transparency width.

Depending on the degree of collectivity and the applied driving fields, the system exhibits two limiting behaviors: 
transient superradiant emission when the probe and control are switched off, 
and collective EIT in the stationary regime under continuous excitation. 
Both effects are captured by a single mean-field master equation derived below, which effectively describes a representative atom subject to collective feedback from the ensemble. 
This approach provides a unified framework that connects cooperative radiation and interference phenomena within the same three-level medium.

\subsection{Representative-atom master equation in the Mean-field approximation}
Starting from the collective master equation for an ensemble of $N$ identical three-level emitters in the Dicke limit,
we derive an effective single-particle equation that incorporates the collective feedback of the ensemble into the dynamics of a representative atom. We start from the collective master equation for an ensemble of $N$ identical three-level atoms in a $\Lambda$ configuration (see Fig.~1), assuming a small and homogeneous sample (Dicke limit) such that all atoms are driven identically and coupled to a common electromagnetic reservoir:

\begin{equation}
\frac{d\rho_N}{dt}
= -i [H_I, \rho_N] + \mathcal{L}^{(31)}\rho_N + \mathcal{L}^{(32)}\rho_N + \mathcal{L}^{(22)}\rho_N + \mathcal{L}^{(33)}\rho_N .
\end{equation}

Here, $\hbar\!=\!1$ and we work in the rotating-wave approximation (RWA). The interaction Hamiltonian is
\begin{equation}
H_I = \Delta_1 S_{33} + (\Delta_1 - \Delta_2) S_{22}
-\tfrac{1}{2}\big[\Omega_p(S_{31}+S_{13}) + \Omega_c(S_{32}+S_{23})\big],
\end{equation}
where we have labeled the single-atom basis as $\ket{1}$, $\ket{2}$ (long-lived/ground states) and $\ket{3}$ (excited state), the single-atom projectors and transition operators as
$\sigma^{(m)}_{ij} = \ket{i}_m\!\bra{j}$,
with $i,j\in\{1,2,3\}$ and $m=1,\dots,N$, and the collective (symmetric) operators as
$S_{ij}=\sum_{m=1}^N \sigma^{(m)}_{ij}$.

The weak \emph{probe} field addresses the $3\!\leftrightarrow\!1$ transition, while the strong \emph{control} field couples the $3\!\leftrightarrow\!2$ transition. 
Their single-atom Rabi frequencies are denoted by $\Omega_p$ (probe) and $\Omega_c$ (control). 
In the Dicke or homogeneous-sample limit, both fields are spatially uniform, with identical amplitude and phase across all atoms. 

We denote  $\omega_{31}$ and $\omega_{32}$ as the bare transition frequencies for $3\!\to\!1$ and $3\!\to\!2$, and $\omega_p$ and $\omega_c$ the probe and control laser frequencies, respectively, and define the \emph{one-photon} detunings as $\Delta_1 \equiv \omega_{31} - \omega_{p},
\Delta_2 \equiv \omega_{32} - \omega_{c}$, and the \emph{two-photon} detuning as $\delta \equiv \Delta_1 - \Delta_2
= (\omega_c-\omega_p) + (\omega_{31}-\omega_{32})$.

For later use, we define generalized ``Pauli'' combinations on each optical branch:
\[
\sigma_x^{(31)} \equiv \sigma_{31}+\sigma_{13},\qquad
i\,\sigma_y^{(31)} \equiv \sigma_{31}-\sigma_{13},
\]
\[
\sigma_x^{(32)} \equiv \sigma_{32}+\sigma_{23},\qquad
i\,\sigma_y^{(32)} \equiv \sigma_{32}-\sigma_{23},
\]
so that the drive part reads $-\tfrac{1}{2}\big(\Omega_p\sigma_x^{(31)}+\Omega_c\sigma_x^{(32)}\big)$ at the single-atom level (and analogously at the collective level with $S_{ij}$).

Assuming Born–Markov–secular approximations---namely, weak system–bath coupling (Born), short bath correlation time (Markov), and neglect of cross terms between nearly degenerate transitions (secular)---the dissipators take the form

\begin{subequations}\label{eq:Lindblad_collective}
\renewcommand{\theequation}{3-\alph{equation}}
\begin{align}
\mathcal{L}^{(31)}\rho_N &= \tfrac{\Gamma_{31}}{2} \big( [S_{13},\rho_N S_{31}] - [S_{31}, S_{13}\rho_N] \big),
\label{eq:L31}\\
\mathcal{L}^{(32)}\rho_N &= \tfrac{\Gamma_{32}}{2} \big( [S_{23},\rho_N S_{32}] - [S_{32}, S_{23}\rho_N] \big),
\label{eq:L32}\\
\mathcal{L}^{(22)}\rho_N &= -\tfrac{\gamma_2}{2} [S_{22}, [S_{22},\rho_N]],
\label{eq:L22}\\
\mathcal{L}^{(33)}\rho_N &= -\tfrac{\gamma_3}{2} [S_{33}, [S_{33},\rho_N]].
\label{eq:L33}
\end{align}
\end{subequations}

We assume permutation symmetry, so that all reduced single-atom states are identical. 
This allows us to trace over $N-1$ atoms and obtain

\begin{subequations}\label{eq:trace_comm}
\renewcommand{\theequation}{4-\alph{equation}}
\begin{align}
\mathrm{Tr}_{2,\dots,N}[S_{\alpha\beta},\rho_N] 
&= [\sigma^{(1)}_{\alpha\beta},\rho],
\qquad
\alpha,\beta\in\{1,2,3\},
\label{eq:trace_comm_a}\\
\mathrm{Tr}_{2,\dots,N}[S_{33},\rho_N]
&= [\sigma^{(1)}_{33},\rho].
\label{eq:trace_comm_b}
\end{align}
\end{subequations}
with analogous relations holding for $S_{22}, S_{31}, S_{13}, S_{32},$ and $S_{23}$ and

\begin{widetext}
\begin{subequations}\label{eq:MF_sums}
\renewcommand{\theequation}{\theparentequation-\alph{equation}}
\begin{align}
\sum_{r=p+1}^N 
\mathrm{Tr}_{p+1,\ldots,N}\,
\sigma_r \rho_N 
&\;\approx\; 
(N-p)\,
\mathrm{Tr}_{p+1,\ldots,N}\,
\sigma_{p+1} \rho_N,
\qquad \text{for } r>p,
\label{eq:MF_sums_a}\\
\sum_{n=2}^N \mathrm{Tr}_{2,\dots,N}\,\sigma^{(n)}_{13}\rho_N 
&\;\approx\; (N-1)\,\langle\sigma_{13}\rangle.
\label{eq:MF_sums_b}
\end{align}
\end{subequations}
\end{widetext}

\noindent
For the radiative channel $3\!\to\!1$, we expand and trace:

\begin{subequations}\label{eq:dissect}
\renewcommand{\theequation}{\theparentequation-\alph{equation}}
\begin{align}
\mathrm{Tr}_{2,\dots,N}[S_{13},\rho_N S_{31}]
&= [\sigma_{13},\rho\sigma_{31}]
  + (N\!-\!1)\,[\langle\sigma_{31}\rangle\sigma_{13},\,\rho],
\label{eq:dissect_a}\\[2pt]
\mathrm{Tr}_{2,\dots,N}[S_{31},S_{13}\rho_N]
&= [\sigma_{31},\sigma_{13}\rho]
  + (N\!-\!1)\,[\langle\sigma_{13}\rangle\sigma_{31},\,\rho],
\label{eq:dissect_b}\\[2pt]
\mathrm{Tr}_{2,\dots,N}L^{(31)}\rho_N
&= \tfrac{\Gamma_{31}}{2}\Big([\sigma_{13},\rho\sigma_{31}] - [\sigma_{31},\sigma_{13}\rho]\Big)\notag\\
&\quad + \tfrac{\Gamma_{31}}{2}(N\!-\!1)\big[\langle\sigma_{31}\rangle\sigma_{13}-\langle\sigma_{13}\rangle\sigma_{31},\,\rho\big],
\label{eq:L31_red}
\end{align}
\end{subequations}

\noindent
The second commutator in \eqref{eq:L31_red} is anti-Hermitian and can be written as a Hamiltonian contribution. With $\sigma_{31}=\tfrac{1}{2}(\sigma_x^{(31)}+i\sigma_y^{(31)})$ and
$\sigma_{13}=\tfrac{1}{2}(\sigma_x^{(31)}-i\sigma_y^{(31)})$, one finds
\begin{equation*}
\begin{aligned}
\big[\langle\sigma_{31}\rangle\sigma_{13} &- \langle\sigma_{13}\rangle\sigma_{31},\,\rho\big] \\
&= -\,i\Big[\tfrac{1}{2}\Big(\langle\sigma_x^{(31)}\rangle\,\sigma_y^{(31)}
      - \langle\sigma_y^{(31)}\rangle\,\sigma_x^{(31)}\Big),\,\rho\Big].
\end{aligned}
\end{equation*}

\noindent
Thus $L^{(31)}$ contributes both a standard single-atom dissipator and a
\emph{collective mean-field} Hamiltonian term $\propto (N-1)$.
An identical reduction holds for the $3\!\to\!2$ channel (replace $31\to32$).
For pure dephasing, the trace gives directly
$\mathrm{Tr}_{2,\dots,N}[S_{aa},[S_{aa},\rho_N]]
=[\sigma_{aa},[\sigma_{aa},\rho]]$ for $a\in\{2,3\}$.
Grouping the terms, we obtain the representative single three-level particle master equation:

\begin{equation}
\frac{d\rho}{dt}
= -i\,[H_{\mathrm{eff}},\rho] + \mathcal{L}_D[\rho],
\label{eq:repME}
\end{equation}

\noindent
where the representative effective Hamiltonian reads
\begin{equation}
\label{eq:Heff}
\begin{aligned}
H_{\mathrm{eff}}
&= \Delta_1\,\sigma_{33} + (\Delta_1-\Delta_2)\,\sigma_{22} \\[2pt]
&\quad - \tfrac{1}{2}\big(\Omega_p\,\sigma_x^{(31)}+\Omega_c\,\sigma_x^{(32)}\big) \\[2pt]
&\quad + \tfrac{\Gamma_{31}}{2}(N-1)\!\left(\langle\sigma_x^{(31)}\rangle\,\sigma_y^{(31)} - \langle\sigma_y^{(31)}\rangle\,\sigma_x^{(31)}\right) \\[2pt]
&\quad + \tfrac{\Gamma_{32}}{2}(N-1)\!\left(\langle\sigma_x^{(32)}\rangle\,\sigma_y^{(32)} - \langle\sigma_y^{(32)}\rangle\,\sigma_x^{(32)}\right) \, ,
\end{aligned}
\end{equation}
and $\mathcal{L}_D[\rho]$ collects the standard single-atom dissipators:

\begin{equation}\label{eq:master_eq1}\
\begin{aligned}
\mathcal{L}_D[\rho]
&= \frac{\Gamma_{31}}{2}\Big([\sigma_{13},\rho\,\sigma_{31}] - [\sigma_{31},\sigma_{13}\rho]\Big) \\[2pt]
&\quad + \frac{\Gamma_{32}}{2}\Big([\sigma_{23},\rho\,\sigma_{32}] - [\sigma_{32},\sigma_{23}\rho]\Big) \\[2pt]
&\quad - \frac{\gamma_2}{2}\,[\sigma_{22},[\sigma_{22},\rho]]
      - \frac{\gamma_3}{2}\,[\sigma_{33},[\sigma_{33},\rho]] \, .
\end{aligned}
\end{equation}

In the exact treatment, for this three-level $\Lambda$ system the collective Hamiltonian contains two radiative channels, $|3\rangle \to |1\rangle$ and $|3\rangle \to |2\rangle$, each of which can undergo superradiance independently, see Fig. 2 and 3. The corresponding collective operators
\begin{equation}
R_{31} = \sum_{\alpha=1}^N |3\rangle_\alpha \langle 1|, 
\qquad 
R_{32} = \sum_{\alpha=1}^N |3\rangle_\alpha \langle 2|,
\end{equation}
mediate the cooperative emission along the two decay paths. 
At short times the expectation values $\langle R_{31}^\dagger R_{31}\rangle$ and $\langle R_{32}^\dagger R_{32}\rangle$ increase as the macroscopic polarization develops, producing the characteristic SR burst in each transition. 

Summarizing, by analyzing Eq.~(13) with pumps turned off, we capture the universal cooperative transient: an initially inverted population decays through collective emission, with $I_{31}(t)\sim \langle R_{31}^\dagger R_{31}\rangle$ and $I_{32}(t)\sim \langle R_{32}^\dagger R_{32}\rangle$ showing superradiant power scaling. 
This provides the natural benchmark to later connect with the stationary interference regime (EIT) once the driving fields are restored.

Equations~\eqref{eq:repME}–\eqref{eq:Heff} define the master equation of the representative atom, providing a unified framework that connects the transient Dicke superradiance - through cooperative terms $\propto N{-}1$ - to a stationary collective EIT governed by the interference of the control field.

\subsubsection{Consistency checks and validation against exact dynamics}

To verify the accuracy of the representative-atom master equation, we numerically compared its predictions with the full collective dynamics solved in the symmetric subspace. As shown in the main text [see Fig. 2 and 3], the mean-field (MF) formulation reproduces both the transient and stationary regimes with excellent quantitative agreement. In the superradiant domain, the MF solution captures the cooperative burst intensity, peak time, and scaling exponent $I_{\max}\!\sim\!N^2$ with deviations below a few percent even for moderate $N$. In the stationary regime, the steady-state optical coherences obtained from the MF model yield absorption and dispersion spectra that closely match the exact collective calculation, including the narrowing and contrast reduction of the EIT window. 

In the single-atom limit ($N=1$), the collective feedback terms vanish, and the master equation reduces exactly to the standard optical Bloch equations for a three-level $\Lambda$ system, reproducing the well-known EIT lineshape with perfect transparency at two-photon resonance. This limiting behavior confirms the consistency of the representative-atom framework across the entire range from independent emitters to fully collective ensembles.

Overall, these checks demonstrate that the representative-atom approach faithfully captures the essential $N$-dependent feedback that governs both the superradiant emission and the collective EIT transparency, providing a computationally efficient and physically transparent alternative to the full $N$-body treatment.

\subsection{Superradiance}
We focus on the short-time dynamics generated by the representative-atom master equation in the absence of external pumps. 
For times $t \ll \Gamma^{-1}_{ij}$ the evolution is dominated by the coherent mean-field Hamiltonian, and the dissipative part of the Liouvillian $\mathcal{L}_D$ can be neglected. 
In this regime, both probe and control fields are switched off, such that the dynamics are entirely governed by the initial collective excitation. 

The mean-field Hamiltonian derived in the Dicke limit (see SM) governs the purely coherent evolution of the reduced density matrix.
With the ensemble initially prepared in an inverted configuration, the subsequent evolution describes the buildup of cooperative emission along the two radiative branches, $|3\rangle\!\to\!|1\rangle$ and $|3\rangle\!\to\!|2\rangle$.
The corresponding intensities $I_{31}(t)$ and $I_{32}(t)$ are proportional to the photon fluxes emitted on the two radiative branches,
$I_{3\alpha}(t)\propto \Gamma_{3\alpha}\langle R_{3\alpha}^\dagger R_{3\alpha}\rangle$ ($\alpha=1,2$),
where $R_{3\alpha}=\sum_{n=1}^N \ket{\alpha}_n\!\bra{3}$ is the collective (Dicke) lowering operator (see SM).

\subsection{Symmetric and Asymmetric decay channels}
\begin{figure*}[t]
\centering

\begin{subfigure}[t]{0.52\textwidth}
    \centering
    \includegraphics[width=\linewidth]{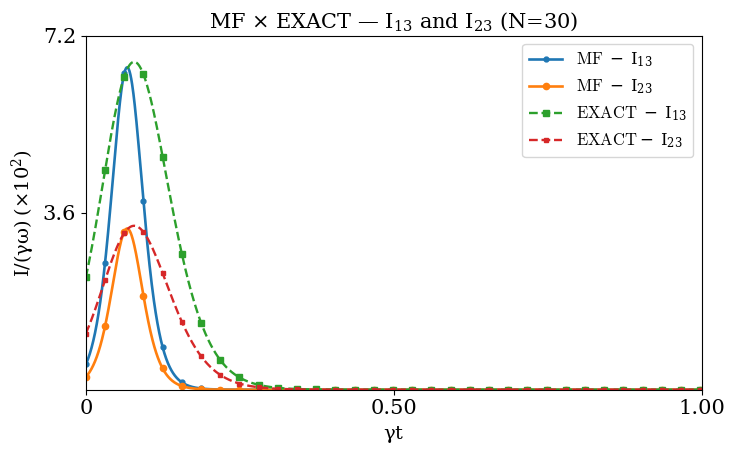}
    \label{fig:transient_asym}
\end{subfigure}
\hfill
\begin{subfigure}[t]{0.46\textwidth}
    \centering
    \includegraphics[width=\linewidth]{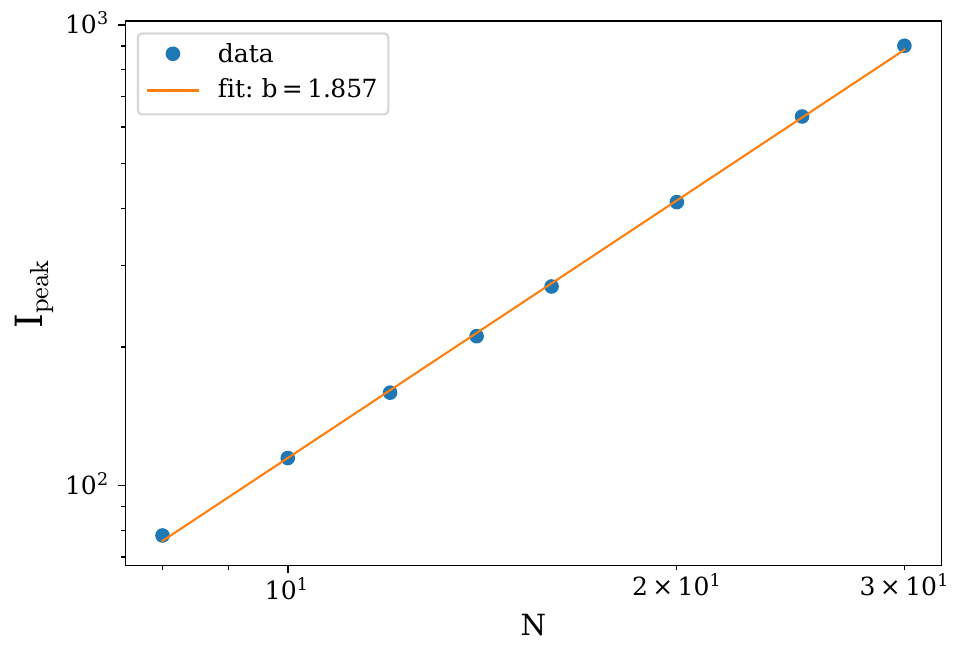}
    \label{fig:scaling_asym}
\end{subfigure}

\caption{Transient superradiant dynamics for asymmetric decay, $\Gamma_{31}=5\Gamma_{32}$.
Left: Channel-resolved intensities $I_{13}(t)$ and $I_{23}(t)$ for $N=30$ with $\Omega_p=\Omega_c=0$ and $\gamma_2=0.01\Gamma$; mean-field (solid) vs exact (dashed).
Right: Peak intensity $I_{\mathrm{peak}}$ versus $N$ with a power-law fit ($b\simeq 1.86$).}
\label{fig:SR_asym_scaling}
\end{figure*}

\begin{figure*}[tb]
\centering

\begin{subfigure}[t]{0.52\textwidth}
    \centering
    \includegraphics[width=\linewidth]{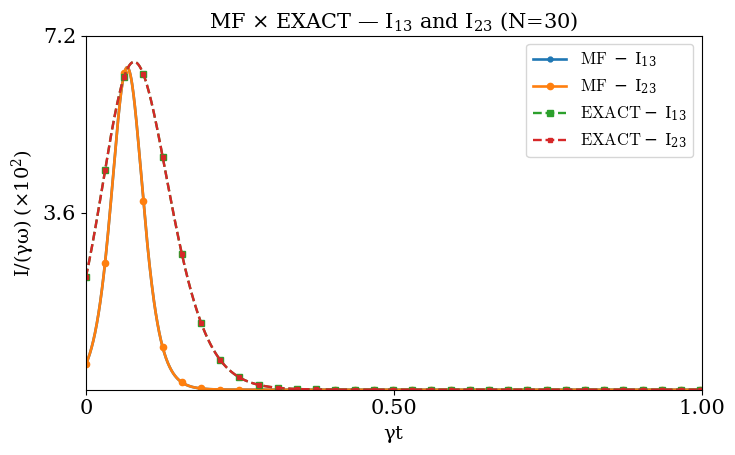}
    \label{fig:transient_sym_SM}
\end{subfigure}
\hfill
\begin{subfigure}[t]{0.45\textwidth}
    \centering
    \includegraphics[width=\linewidth]{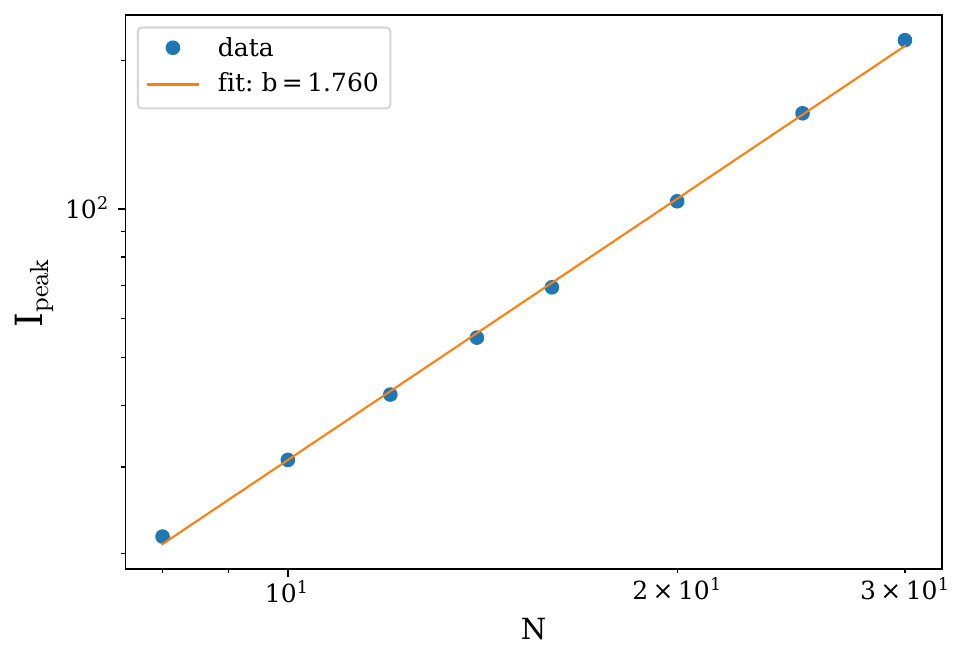}
    \label{fig:scaling_sym_SM}
\end{subfigure}

\caption{Transient superradiant dynamics (symmetric decay channels).
Left: channel-resolved emission for $N=30$ atoms with equal decay rates ($\Gamma_{31}=\Gamma_{32}=\Gamma$), $\gamma_2=0.01\Gamma$, and splittings
$\omega_{31}=\omega$, $\omega_{32}=0.5\,\omega$ ($\omega/\Gamma=1000$), obtained in the absence of coherent drives
($\Omega_p=\Omega_c=0$) from the initial state
$\ket{\psi(0)}=\sqrt{1-2\varepsilon^2}\ket{3}+\varepsilon\ket{1}+\varepsilon\ket{2}$ with $\varepsilon=0.1$.
Solid lines with circular markers correspond to the mean-field result, while dashed lines with square markers denote the exact dynamics in the symmetric subspace.
Right: scaling of the peak intensity $I_{\text{peak}}$ with atom number $N$, where a log--log fit yields
$I_{\text{peak}}\sim N^{b}$ with $b\approx1.76$, confirming collective enhancement with finite-size corrections.}
\label{fig:SR_equal_scaling_SM}
\end{figure*}

When the two radiative channels differ in strength, e.g., $\Gamma_{31}=5\,\Gamma_{32}$, emission through the $3\!\to\!1$ branch dominates.
Simulations are performed in the rotating frame with no coherent drives ($\Omega_p=\Omega_c=0$), using the same parameters as in the symmetric case, except for the decay rates.
As shown in Fig.~\ref{fig:SR_asym_scaling} (left), the $3\!\to\!1$ channel produces a sharper and more intense burst at $\Gamma t_{\text{peak}}\!\approx\!0.03$, while $I_{32}(t)$ remains comparatively weaker.
The total emission $I_{\mathrm{tot}}(t)=I_{31}(t)+I_{32}(t)$ inherits this asymmetry and exhibits a pronounced cooperative peak.
The scaling analysis in Fig.~\ref{fig:SR_asym_scaling} (right) yields $I_{\text{peak}}\sim N^{b}$ with $b\!\approx\!1.86$, indicating that channel imbalance enhances the effective cooperative buildup. The symmetric case, $\Gamma_{31}=\Gamma_{32}$, is shown in Fig.~\ref{fig:SR_equal_scaling_SM}, confirming the same trends and comparable mean-field/exact agreement.

\subsection{Apparent exponents and finite-size bias}
Finite ensembles exhibit apparent exponents $\xi(N)$ that deviate from the ideal $N^2$ law due to constant prefactor effects. Here we provide the corresponding derivation, and we also
emphasize an important methodological point: the identification of the
$1/\ln N$ finite-size correction for the apparent exponent $\xi(N)$ relies on
the ability to \emph{infer the functional form of the burst} from a direct
benchmark against the numerically exact dynamics in the symmetric subspace.
In other words, adopting a standard superradiant envelope is not an \emph{a
priori} assumption in our work; it is validated by comparison with the exact
time trace, which in turn makes the extraction of $I_{\max}(N)$ and the
analysis of $\xi(N)$ well-defined and robust.

With both driving fields switched off ($\Omega_p=\Omega_c=0$), the collective
radiative channels are governed by the jump operators $R_{31}$ and $R_{32}$
[cf.\ Eq.~(15)]. The (unnormalized) emitted intensities scale as
\begin{equation}
I_{31}(t)\propto \langle R_{31}^\dagger R_{31}\rangle,\qquad
I_{32}(t)\propto \langle R_{32}^\dagger R_{32}\rangle .
\label{eq:I_def}
\end{equation}
In the Dicke limit, $R_{3i}$ is a collective operator whose matrix elements
scale linearly with $N$, so the peak intensity can scale quadratically,
$I_{\max}(N)\propto N^2$ in the cooperative-emission regime.

In practice, we extract the peak value $I_{\max}(N)$ from the numerical burst.
Crucially, the \emph{exact} simulations allow us to infer (and explicitly
verify) that, in the vicinity of the maximum, the temporal profile is well
captured by the standard superradiant envelope
\begin{equation}
I(t)\simeq I_{\max}(N)\,\mathrm{sech}^2\!\left(\frac{t-t_d}{\tau}\right),
\label{eq:sech2_fit}
\end{equation}
where $t_d$ is the delay time and $\tau$ is the burst duration. This
benchmarking step is what enables a controlled definition of the peak
intensity and, consequently, of the apparent exponent $\xi(N)$: without the
comparison to the exact dynamics, the use of a specific envelope (and thus
the very meaning of the extracted $I_{\max}(N)$ in finite-size samples) would
remain model-dependent.

For fixed microscopic parameters and a fixed initial preparation,
the peak intensity can be written as
\begin{equation}
I_{\max}(N)= I_0\,N^2\,A ,
\label{eq:Imax_scaling}
\end{equation}
where $I_0$ is an $N$-independent single-emitter scale and $A>0$ is an
$N$-independent dimensionless prefactor (set by the initial condition and
branching details, but not by $N$).

We define the apparent exponent $\xi(N)$ through
\begin{equation}
I_{\max}(N)= I_0\,N^{\xi(N)} \quad\Longrightarrow\quad
\xi(N)=\frac{\ln\!\big[I_{\max}(N)/I_0\big]}{\ln N}.
\label{eq:xi_def}
\end{equation}
Using Eq.~\eqref{eq:Imax_scaling}, we obtain
\begin{equation}
\xi(N)=\frac{\ln\!\big(N^2A\big)}{\ln N}
=2+\frac{\ln A}{\ln N}.
\label{eq:xi_result}
\end{equation}
Therefore, the asymptotic superradiant exponent is $\lim_{N\to\infty}\xi(N)=2$,
and the finite-size correction obeys
\begin{equation}
|\xi(N)-2|=\frac{|\ln A|}{\ln N}\sim \frac{1}{\ln N}.
\label{eq:xi_correction}
\end{equation}
We note that this logarithmic correction assumes that $A$ does not scale with $N$, as is the case in our model. If the prefactor acquired an $N$-dependence (e.g., $A\propto N^\alpha$), the effective exponent would be shifted by $\alpha$. The deviation produces values slightly below or above $2$ for moderate $N$.  
This explains why the extracted slopes in Fig.~\ref{fig:SR_asym_scaling} and Fig.~\ref{fig:SR_equal_scaling_SM}, fall around $b\approx 1.7-1.8$,
converging to the universal quadratic scaling as $N$ increases.

\section{Asymptotic steady-state regime} When the drives are turned on, in the asymptotic limit $d\rho/dt = 0$, the dynamics reduce to a set of stationary equations for the optical coherences. 
In the linear regime (weak probe, $\rho_{33}\!\approx\!0$, $\rho_{11}\!\approx\!1$), the relevant steady-state equations are obtained from the mean-field master equation Eq.~\ref{eq:repME}, yielding, for the representative three-level atom,
\begin{equation}
\rho_{31} = 
\frac{ i \, \dfrac{\Omega_p}{2} \left( \tfrac{\gamma_{2}}{2} + i(\Delta_1 - \Delta_2) \right) }
{ \big( \Gamma^{(\mathrm{eff})} + i\Delta_1 \big)
  \left( \tfrac{\gamma_{2}}{2} + i(\Delta_1 - \Delta_2) \right)
  + \tfrac{\Omega_c^2}{4} },
\label{eq:ss_s31}
\end{equation}
where $\Gamma^{(\mathrm{eff})}
= (\Gamma_{31} + \Gamma_{32})/2
+ \gamma_3
+ \Gamma_{31}(N{-}1)$. Since Eq.~\eqref{eq:ss_s31} refers to a representative atom, 
the corresponding collective polarization must be multiplied by $N$ 
to compare with the full ensemble response.

Equation~\eqref{eq:ss_s31} highlights three key ingredients: 
(i) the effective excited-state width $\Gamma^{(\mathrm{eff})}$, which incorporates natural decay $\tfrac{\Gamma_{31}+\Gamma_{32}}{2}$, pure dephasing $\gamma_{3}$, and the collective broadening term $\Gamma_{31}(N{-}1)$—this effective width sets the overall absorption scale; 
(ii) the ground-state dephasing $\gamma_{2}$, which fills the transparency dip and limits the contrast of the EIT window; and 
(iii) the control-field Rabi frequency $\Omega_{c}$, which induces Autler–Townes splitting and opens the transparency window. 
A well-resolved EIT window requires that the control-induced splitting exceed the ground-state dephasing, 
$\Omega_{c}^{2} \gtrsim 4\,\Gamma^{(\mathrm{eff})}\gamma_{2}$.

\begin{figure}[t!]
  \centering
  \includegraphics[width=\columnwidth]{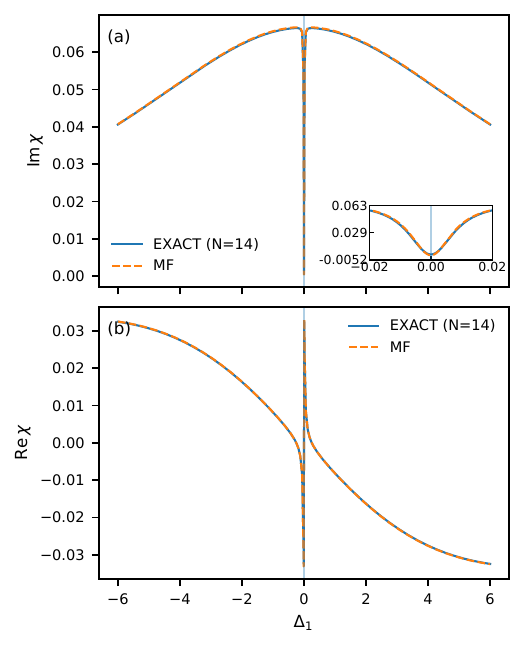}
  \caption{Exact (solid) vs mean-field (dashed) susceptibility for $N=14$:
  (a) absorption $\mathrm{Im}\,\chi(\Delta_1)$ and (b) dispersion $\mathrm{Re}\,\chi(\Delta_1)$.
  The inset in panel (a) shows a magnified view of the narrow EIT transparency feature around $\Delta_1\simeq 0$.}
  \label{fig:EIT_full}
\end{figure}

Figure~\ref{fig:EIT_full}(a)-(b) illustrates the comparison between the exact collective calculation (solid black) and the mean-field approximation (dashed blue) for a finite ensemble ($N=14$). 
Panel~(a) shows the probe absorption, proportional to the imaginary part of the susceptibility $\mathrm{Im}\,\chi(\Delta_1)$, 
while panel~(b) displays the corresponding dispersion, given by the real part $\mathrm{Re}\,\chi(\Delta_1)$. The parameters used are probe Rabi frequency $\Omega_p = 0.1\Gamma$, control Rabi frequency $\Omega_c = 0.5\Gamma$, equal decay rates $\Gamma_{31}=\Gamma_{32}=\Gamma$, vanishing control detuning $\Delta_2=0$, and a very small dephasing $\gamma_{2}=\gamma_{3}=10^{-4}\Gamma$.
Note the excellent quantitative agreement is observed between the two approaches, with only minor deviations near resonance.
In particular, the transparency dip at $\Delta_1=0$ becomes noticeably narrower, highlighting how collective broadening modifies the EIT window. 
This consistency confirms that the representative-atom model faithfully captures the essential $N$-dependent scaling of the full collective dynamics.

The narrowing of the transparency window has important practical implications: it acts as an ultra-narrow spectral filter and enables high-resolution frequency discrimination.
This feature is widely exploited in EIT-based metrology, where the sharp transparency dip allows for the precise determination of resonance shifts (e.g., in atomic clocks or magnetometry), and in slow-light and quantum-memory protocols, where the enhanced dispersion enables the efficient storage of narrowband optical pulses
\cite{vanier2005atomic,shahriar2007ultrahigh,budker2007optical,savukov2005nmr,fleischhauer2000dark,fleischhauer2005electromagnetically,hau1999light,hammerer2010quantum,macovei2005coherent, ariunbold2025superradiant}.

\section{Analytic scaling of the EIT window with $N$}
In the weak–probe, steady–state regime with $\Delta_2=0$, the normalized linear susceptibility of a $\Lambda$ system under a control field $\Omega_c$ follows Eq.~(\ref{eq:ss_s31}), where $\Omega_p$ and $\Omega_c$ are the Rabi frequencies of the weak probe and strong control fields, respectively;  
$\Delta_1$ and $\Delta_2$ denote the probe and control detunings;  
$\gamma_2$ is the ground-state dephasing rate;  
$\gamma_3$ accounts for the excited-state pure dephasing;  
$\Gamma_{31}$ and $\Gamma_{32}$ are the spontaneous emission rates from the excited level $|3\rangle$ to the lower states $|1\rangle$ and $|2\rangle$;  
and $N$ is the number of emitters in the ensemble, entering through the collective broadening term $\Gamma_{31}(N{-}1)$ that defines the effective linewidth $\Gamma^{(\mathrm{eff})}$.

The single-atom susceptibility is defined as $\tilde{\chi} = \rho_{31}/\Omega_p$, and the macroscopic susceptibility of the medium as $\chi = C\,\tilde{\chi}$$, C = n_{\!at}|\mu_{31}|^2 \varepsilon_0\hbar$, where $n_{\!at}$ is the atomic density, $\mu_{31}$ the dipole moment of the optical transition,  
$\varepsilon_0$ the vacuum permittivity, and $\hbar$ the reduced Planck constant.

Separating real and imaginary parts yields
\begin{subequations}\label{eq:chi}
\renewcommand{\theequation}{\theparentequation-\alph{equation}}
\begin{align}
\mathrm{Im}\,\chi(\Delta_1,\Delta_2)
&= C\,\mathrm{Im}\!\left(\frac{\rho_{31}}{\Omega_p}\right), \label{eq:24a}\\[3pt]
\mathrm{Re}\,\chi(\Delta_1,\Delta_2)
&= C\,\mathrm{Re}\!\left(\frac{\rho_{31}}{\Omega_p}\right), \label{eq:24b}
\end{align}
\end{subequations}
which represent, respectively, the absorptive and dispersive responses of the medium.


\subsection{Consistency check: susceptibility from various solvers}
\label{sec:chi_consistency}

\begin{figure*}[t]
  \centering
  \includegraphics[width=\textwidth]{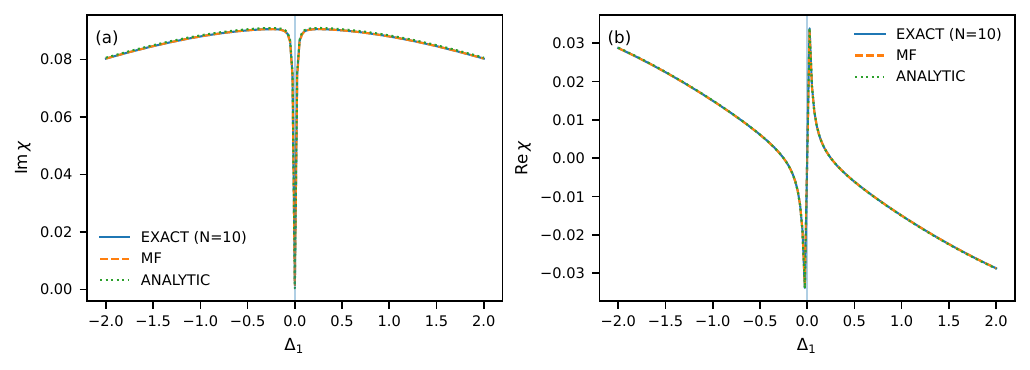}
  \caption{\textbf{Susceptibility consistency check across solvers.}
  Comparison of the steady-state probe susceptibility obtained from the
  exact collective solver (symmetric subspace), the representative-atom master equation
  (steady state), and the analytic linear-probe expression, as indicated in the legend.
  (a) Absorption plotted as a positive quantity, $-\mathrm{Im}\,\chi(\Delta_1)$.
  (b) Dispersion, $\mathrm{Re}\,\chi(\Delta_1)$.
  The detuning is shown in units of $\Gamma_{31}$, and the susceptibility is normalized
  per emitter (see text).}
  \label{fig:chi_exact_rep_analytic}
\end{figure*}

As an explicit numerical consistency check, we compare the steady-state probe
susceptibility computed with three independent approaches:
(i) the \emph{exact collective} steady-state solver in the symmetric (Dicke) subspace,
yielding the per-emitter susceptibility
$\chi_{\mathrm{exact}}(\Delta_1)=\langle S_{1}\rangle_{\mathrm{ss}}/(N\Omega_p)$;
(ii) the \emph{representative-atom master equation} (representative ME) solved numerically
for its steady state, from which we extract
$\chi_{\mathrm{rep}}(\Delta_1)=\langle\sigma_{13}\rangle_{\mathrm{ss}}/\Omega_p$; and
(iii) the \emph{analytic linear-probe} closed-form expression for
$\chi_{\mathrm{an}}(\Delta_1)=\rho_{31}^{(\mathrm{lin})}(\Delta_1)/\Omega_p$.
In Fig.~\ref{fig:chi_exact_rep_analytic} we plot the corresponding absorption
$-\mathrm{Im}\,\chi$ and dispersion $\mathrm{Re}\,\chi$ as functions of the normalized detuning
$\Delta_1/\Gamma_{31}$ (all quantities are dimensionless in units of $\Gamma_{31}$).
The representative-ME curve is found to coincide with the analytic linear-probe
expression (within numerical tolerance), and both track the exact collective
result over the entire scan range; in practice, the representative-ME and analytic
curves often overlap.

\subsection{EIT window and contrast}

The optical susceptibility of the probe transition to the sample is proportional to the steady-state coherence normalized by the probe Rabi frequency:

\begin{widetext}
\begin{equation}
\mathfrak{I}\!\left(\frac{N\rho_{31}}{\Omega_p}\right) =
\frac{
N\Gamma^{(\mathrm{eff})}\!\left(\tfrac{\gamma_{2}^{2}}{4}+(\Delta_{1}-\Delta_{2})^{2}\right)
+\tfrac{\gamma_{2}\Omega_{c}^{2}}{4}
}{
2\Big[
\big(\Gamma^{(\mathrm{eff})}\gamma_{2}-\Delta_{1}(\Delta_{1}-\Delta_{2})+\tfrac{\Omega_{c}^{2}}{4}\big)^{2}
+\Gamma^{(\mathrm{eff})}
(\Delta_{1}-\Delta_{2})+\gamma_{2}\Delta_{1}\big)^{2}
\Big]
},
\label{eq:Im_rho31}
\end{equation}
\end{widetext}
where $\Gamma^\mathrm{(eff)}=(\Gamma_{31} + \Gamma_{32})/2 + \gamma_{3} + \Gamma_{31}(N-1)$.

Eq.~(\ref{eq:Im_rho31}) shows that the number $N$ of two-level systems affects the susceptibility in two distinct ways. First, in the representative--atom formulation the macroscopic polarization is given by 
$P \propto N\,\langle\rho_{31}\rangle_{\mathrm{ss}}$, so that the ensemble susceptibility carries an overall factor of $N$. 
Second, the denominator of Eq.~(\ref{eq:Im_rho31}) contains the collective broadening 
$\Gamma_{31}(N{-}1)$, which scales linearly with $N$ and modifies the lineshape itself. Near two-photon resonance

\begin{subequations}\label{eq:SM_EITwidth}
\renewcommand{\theequation}{\theparentequation-\alph{equation}}
\begin{align}
\Delta_{\mathrm{EIT}}(N) &\approx \gamma_2+\frac{\Omega_c^2}{\Gamma^{(\mathrm{eff})}(N)},
\label{eq:SM_EITwidth_a}\\
\Gamma^{(\mathrm{eff})}(N) &\simeq \Gamma_{31}N\ (N\gg1)\ \Rightarrow\
\Delta_{\mathrm{EIT}} \simeq \gamma_2+\frac{\Omega_c^2}{\Gamma_{31}}\frac{1}{N}.
\label{eq:SM_EITwidth_b}
\end{align}
\end{subequations}

with the window resolved condition: \(\Omega_c^2 \gtrsim 4\,\Gamma^{(\mathrm{eff})}\gamma_2\). 
The physical consequences of the collective effects are therefore twofold:  
(i) the overall amplitude of the polarization grows linearly with $N$, which enhances the light--matter coupling strength of the medium;  
(ii) the effective excited-state linewidth broadens with $N$, which reduces the dip contrast, while \emph{the EIT transparency window itself narrows with $N$} - see Fig.\ref{fig:EIT_full}.

\,

\subsection{Group velocity} From the real part of Eq.~\eqref{eq:ss_s31} we obtain the dispersive response of the sample, and hence the group velocity. With the control field on resonance \((\Delta_2=0)\), the probe dispersion becomes
\begin{widetext}
\begin{equation}
\chi_{R}=\frac{n_{at}|\mu_{31}|^2}{\hbar\varepsilon_0}\,\frac{\Omega_p}{2}\,
\frac{\Delta_{1}\!\Big(\tfrac{\gamma_{2}^{2}}{4}+\Delta_{1}^{2}\Big)
-\tfrac{\Omega_{c}^{2}}{4}\,\Delta_{1}}
{\Big(\Gamma^{(\mathrm{eff})}\tfrac{\gamma_{2}}{2}
-\Delta_{1}^{2}
+\tfrac{\Omega_{c}^{2}}{4}\Big)^{2}
+\Big[\Delta_{1}\Big(\Gamma^{(\mathrm{eff})}+\tfrac{\gamma_{2}}{2}\Big)\Big]^{2}} .
\label{eq:Re_rho31_D2zero}
\end{equation}
\end{widetext}

The group velocity is related to the refraction index $n(\omega,\Omega)$ as
\begin{equation}
    v_g(\omega,\Omega) \;=\; \frac{c}{\,n(\omega,\Omega) + \omega \,\tfrac{dn}{d\omega}} \,,
\qquad 
n(\omega,\Omega) = \sqrt{\,1 + \chi_R}\,
\end{equation}
and can be obtained analytically as follows:
\begin{equation}
v_g(N) \;=\; 
\frac{c}{\,1 + \dfrac{\omega_p}{2}\,C\,
\left.\frac{d}{d\Delta_1}\,\mathrm{Re}\,\rho_{31}\right|_{\Delta_1=0}} ,
\end{equation}
where the line--center slope driving the group index is
\begin{equation}
\left.\frac{d}{d\Delta_1}\mathrm{Re}\!\big(N\rho_{31}\big)\right|_{\Delta_1=0}
=\frac{N\Omega_p}{2}\,
\frac{ \big(\tfrac{\gamma_2}{2}\big)^{2} - \big(\tfrac{\Omega_c}{2}\big)^{2} }{[A_0(N)]^2},
\label{eq:S0_D2zero}
\end{equation}
and we have defined
\begin{equation}
A_0(N)=\tfrac{\gamma_2}{2}\,\Gamma^{\mathrm{(eff)}}(N)+\tfrac{\Omega_c^2}{4}.
\label{eq:A0_def}
\end{equation}

The group index becomes
\begin{equation}
n_g\simeq 1+\frac{\omega_p C}{4}\,
\frac{ \big(\tfrac{\gamma_2}{2}\big)^{2} - \big(\tfrac{\Omega_c}{2}\big)^{2} }{A_0(N)^2},
\label{eq:ng_D2zero}
\end{equation}
and thus the group velocity is
\begin{equation}
\frac{v_g}{c}\simeq\frac{1}{1+\delta(N)} ,
\label{eq:vg_D2zero}
\end{equation}
with
\begin{equation}
\delta(N)=\frac{\omega_p C}{4}\,
\frac{ \big(\tfrac{\gamma_2}{2}\big)^{2} - \big(\tfrac{\Omega_c}{2}\big)^{2} }
{ \bigl(\tfrac{\gamma_2}{2}\Gamma^{\mathrm{(eff)}}(N)+\tfrac{\Omega_c^2}{4}\bigr)^{2}} .
\label{eq:deltaN_D2zero}
\end{equation}

Equations~\eqref{eq:ng_D2zero}–\eqref{eq:deltaN_D2zero} show that the $N$--dependence of $v_g$ enters only through $A_0(N)$ and $\Gamma^{(\mathrm{eff})}(N)$.

For large $N$ one has $A_0(N)\simeq \tfrac{\gamma_2}{2}\Gamma_{31}N$, hence
\begin{equation}
    \delta(N)\;\simeq\;\frac{\omega_p C}{4}\,
\frac{\big(\tfrac{\gamma_2}{2}\big)^2-\big(\tfrac{\Omega_c}{2}\big)^2}{\big(\tfrac{\gamma_2}{2}\Gamma_{31}N\big)^2},
\label{partcular}
\end{equation}
such that
\begin{equation}
    \frac{v_g}{c}\;\simeq\;\frac{1}{1+\delta(N)}\;=\;1-\operatorname{sgn}\!\Big(\frac{\Omega_c^2}{\gamma_2^2}-1\Big)\,\frac{|\!K\!|}{N^2}+o(N^{-2}),
\end{equation}
with
\begin{equation}
    K=\frac{\omega_p C}{4\Gamma_{31}^2}\left|1-\frac{\Omega_c^2}{\gamma_2^2}\right|.
\end{equation}

\begin{figure}[t]
\centering
\includegraphics[width=0.7\linewidth]{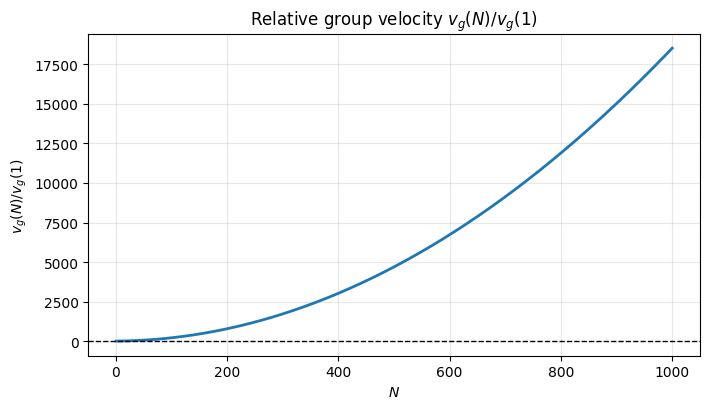}
\caption{Relative group velocity $v_g(N)/v_g(1)$ as a function of the number of emitters $N$. 
The result shows that, although the absolute group velocity remains well below $c$, it increases strongly with $N$, scaling approximately as $N^2$ in the large-$N$ limit. 
This highlights the counterintuitive effect that cooperative broadening makes the medium more transparent, leading to faster propagation compared to the single-atom case.}
\label{fig:vg_ratio}
\end{figure}
When considering the strong slow-light regime ($|\delta|\gg 1$), the cooperative broadening $\Gamma_{\mathrm{eff}}(N)\propto N$ fully determines the scaling. In this limit, the absolute group velocity $v_g(N)$ saturates to a constant value (well below $c$ due to the prefactor $C$), with corrections vanishing as $1/N^2$. Therefore, in the strong slow-light regime the quadratic $N^2$ scaling emerges clearly, since $v_g(1)$ is much smaller than the asymptotic value, being strongly suppressed by cooperative broadening,  a highly counterintuitive effect. 
Note that for $\Omega_c<\gamma_2$ (normal dispersion), one has $\delta>0$ and $v_g<c$, while for $\Omega_c>\gamma_2$ (anomalous dispersion), $\delta<0$ and $v_g>c$. 
In both cases, $v_g$ changes with $N$ but remains far from $c$ due to the scaling factor $C$. Fig.~\ref{fig:vg_ratio} shows how the ratio $v_g(N)/v_g(1)$ scales with $N$ taking into account $\delta(N)$ as given by Eq.\eqref{eq:deltaN_D2zero}.

For large $N$, we find the striking ratio $v_g(N)/v_g(1) \approx N^2$, where $v_g(1)$ denotes the conventional, non-collective EIT group velocity for a single atom. Although collective effects are generally expected to enhance optical slowing, this result is highly counterintuitive: the cooperative coupling instead increases the \emph{relative} propagation speed when compared to the single-atom case. At the same time, the absolute group velocity $v_g(N)$ remains well below $c$ and continues to be constrained by optical depth and decoherence, in agreement with the general slow-light limitations discussed in Ref.~[23], as reflected in the scaling factors presented in the SM.
To see this, we adopt \(N=300\) as a realistic operating point under the sodium D\(_2\) parameters \(\lambda_p = 589~\mathrm{nm}\), \(n_{\mathrm{at}} = 10^{20}~\mathrm{m^{-3}}\), \(|\mu_{31}| = 3.0\times10^{-29}~\mathrm{C\,m}\), \(\gamma_2/2\pi = 0.5~\mathrm{kHz}\), \(\Gamma_{31}/2\pi = \Gamma_{32}/2\pi = 5~\mathrm{MHz}\), \(\gamma_3 \simeq 0\), \(\Omega_c/2\pi = 1.5~\mathrm{MHz}\), and \(\Omega_p = 0.05~\mathrm{rad/s}\) in the weak-probe regime. The choice \(N=300\) guarantees the EIT consistency condition that the effective broadening remains below the control-induced transparency window, given by \(\Gamma_{\mathrm{eff}}(N) \lesssim \Omega_c^2/(4\gamma_2)\), with \(\Gamma_{\mathrm{eff}}(N) \approx \Gamma_{31}N\). In frequency units, this yields \(\Gamma_{\mathrm{eff}}(300) \approx (5~\mathrm{MHz}) \times 300 = 1.5~\mathrm{GHz}\), while \(\Omega_c^2/(4\gamma_2) \approx (1.5~\mathrm{MHz})^2/(2\times0.5~\mathrm{kHz}) \approx 2.25~\mathrm{GHz}\), so the inequality is satisfied and the transparency is preserved. Within the mean-field scaling {$v_g(N)/v_g(1) \approx N^2$, and taking the single-emitter baseline $v_g(1)$ \(\sim 17~\mathrm{m/s}\) as reported by Hau \emph{et al.} \cite{hau1999light}, we obtain \(v_g(300) \approx 300^2 v_g(1) \simeq 9\times10^4\times17~\mathrm{m/s} \approx 1.5\times10^{6}~\mathrm{m/s} \approx 5.1\times10^{-3}c\). This value remains well below the speed of light while exhibiting a strong collective speed-up, showing that \(N=300\) provides a realistic compromise between maintaining EIT transparency and achieving a significant enhancement of the group velocity.

\subsection{Operational test and observable signatures}
A direct experimental test of our collective-broadening mechanism is to measure either
(i) the probe pulse delay $\tau_d(N)=L/v_g(N)$ for a fixed medium length $L$, or
(ii) the line-center dispersion slope $\partial_{\Delta_1}\mathrm{Re}\,\chi|_{\Delta_1=0}$ extracted from phase-sensitive
heterodyne measurements, while varying the effective atom number (or density) at fixed control power.
Our theory predicts a correlated trend: the EIT window narrows with $N$ (Eqs.\ref{eq:SM_EITwidth_a} and \ref{eq:SM_EITwidth_b}) and the relative group
velocity increases as $v_g(N)/v_g(1)\sim N^2$ in the large-$N$ regime, providing two independent, experimentally
accessible signatures of cooperative radiative broadening.

These scaling laws also enable sensitive spectroscopic extraction of the \emph{collective radiative coupling}---mediated by the common radiation field---(e.g., an effective optical depth or the collective broadening entering $\Gamma_{\mathrm{eff}}$) from the EIT window and dispersion.

We thus identify a cooperative limitation to slow light in radiatively coupled $\Lambda$ ensembles: while EIT persists, collective radiative broadening narrows the transparency window (Eq.~\ref{eq:SM_EITwidth_b}) and, in the slow-light regime, yields a pronounced speed-up $v_g(N)/v_g(1)\approx N^2$. Our representative-atom framework therefore provides a practical design rule---tuning $N$ (or the effective density) optimizes the trade-off between transparency bandwidth and delay---and is directly relevant to platforms where the Dicke limit is naturally realized.
Finally, extending the present representative-atom framework to explicitly incorporate inhomogeneous broadening and motional effects in warm-vapor and moving cold-atom ensembles is an interesting direction for future work.

\begin{acknowledgments}
We acknowledge financial support from the Brazilian agencies: Coordenação de Aperfeiçoamento de Pessoal de Nível Superior (CAPES), financial code 001 and CNPq - Conselho Nacional de Desenvolvimento e Pesquisa, Grant 304028/2023-1. NGA and MHYM thank FAPESP Grants 2024/21707-0 and 2024/13689-1.
\end{acknowledgments}

\section*{Appendix: Numerical methods and validation}

This section outlines, step by step, how we build the two solvers used throughout the paper: (i) an \emph{exact} collective solver in the permutation–symmetric subspace, and (ii) a \emph{mean-field} (MF) solver based on a closed-form susceptibility. We also specify the parameter sets and the metrics used to quantify agreement.

\subsection{Exact collective solver}
\label{sec:numerics_exact}

\paragraph*{Step 1 — Symmetric basis.}
We restrict dynamics to the fully symmetric Dicke manifold by labeling basis states as
\(|n_1,n_2,n_e\rangle\) with \(n_1+n_2+n_e=N\).
The resulting Hilbert-space dimension is
\[
D=\#\{(n_1,n_2)\}=\frac{(N+1)(N+2)}{2}.
\]
This reduction preserves all collective observables and enables exact steady-state solutions for moderate \(N\).

\paragraph*{Step 2 — Collective operators.}
We construct diagonal number operators \(N_1,N_2,N_e\) and collective jump operators for the two optical branches,
\[
S_1:~|n_1,n_2,n_e\rangle \mapsto \sqrt{(n_1+1)n_e}\,|n_1{+}1,n_2,n_e{-}1\rangle,
\]
\[
S_2:~|n_1,n_2,n_e\rangle \mapsto \sqrt{(n_2+1)n_e}\,|n_1,n_2{+}1,n_e{-}1\rangle.
\]
We also define the quadratures \(S_{1x}=S_1+S_1^\dagger\) and \(S_{2x}=S_2+S_2^\dagger\).

\paragraph*{Step 3 — Hamiltonian.}
Using the sign convention aligned with the exact code,
\[
H(\Delta_1,\Delta_2)
= \Delta_1\,N_e + (\Delta_1{-}\Delta_2)\,N_2
  + \frac{\Omega_p}{2}S_{1x} + \frac{\Omega_c}{2}S_{2x}.
\]
This form implements one-photon detunings on the excited and \(|2\rangle\) manifolds and coherent drives on the two optical branches.

\paragraph*{Step 4 — Collective dissipators.}
We include collective spontaneous emission and Raman dephasing via collapse operators
\[
C_1=\sqrt{\Gamma_{31}}\,S_1,\qquad
C_2=\sqrt{\Gamma_{32}}\,S_2,\qquad
C_\phi=\sqrt{\gamma_\phi}\,(N_1-N_2).
\]
These Lindblad terms generate radiative decay along \(3{\to}1\) and \(3{\to}2\) and pure dephasing on the ground-state manifold.

\paragraph*{Step 5 — Steady-state scan and susceptibility.}
For each probe detuning \(\Delta_1\) on a uniform grid, we solve
\(\mathcal{L}\rho_{\rm ss}=0\) (QuTiP \texttt{steadystate} with RCM reordering and a direct sparse solver), then compute the probe susceptibility from the collective polarization:
\[
\chi_{\rm exact}(\Delta_1)
=\frac{\langle S_1\rangle_{\rm ss}}{N\,\Omega_p}
=\frac{{\rm Tr}[S_1\rho_{\rm ss}]}{N\,\Omega_p}.
\]
We plot \(\mathrm{Im}\,\chi\) (absorption) and \(\mathrm{Re}\,\chi\) (dispersion). 

\paragraph*{Trace, positivity, and tolerances.}
We verify numerical consistency by enforcing \(|{\rm Tr}\,\rho_{\rm ss}-1|<10^{-12}\) and \(\lambda_{\min}(\rho_{\rm ss})\ge -10^{-10}\) (numerical positivity).
If needed for larger \(D\), we switch to iterative steady-state solvers (BiCGSTAB/GMRES) with tight residual tolerance \(<10^{-10}\).

\paragraph*{Complexity.}
Operator construction is \(\mathcal{O}(D)\) nonzero assignments; steady-state solving is the bottleneck and scales with the sparsity pattern of \(\mathcal{L}\) (typically a few million nonzeros for \(N\sim 20\)–\(30\)).

\subsection{Mean-field solver}
\label{sec:numerics_mf}

\paragraph*{Step A — Effective width and sign alignment.}
The MF susceptibility used in code reads
\[
\chi_{\rm MF}(\Delta_1)
=\frac{i/2}{\Gamma_{\rm eff}(N)+i\Delta_1^{\rm eff}
 + \dfrac{\Omega_c^2}{4\,[\gamma_{2}/2+i(\Delta_1^{\rm eff}-\Delta_2^{\rm eff})]}}\!,
\]
with a sign alignment \(\Delta_1^{\rm eff}=-\Delta_1\) and \((\Delta_1^{\rm eff}-\Delta_2^{\rm eff})=-(\Delta_1-\Delta_2)\) to match the exact convention.
The effective width adopts the collective broadening used in the script,
\[
\Gamma^{(\mathrm{eff})}
= \frac{\Gamma_{31} + \Gamma_{32}}{2}
+ \gamma_3
+ \Gamma_{31}(N{-}1),
\]
which captures the linear-in-\(N\) broadening of the \(3{\to}1\) branch (a more general form with \(\gamma_3\) can be used without changing conclusions).

\paragraph*{Step B — Linear-response normalization.}
Since \(\chi\propto \langle S_1\rangle/(\!N\Omega_p)\), MF curves are independent of \(\Omega_p\) in the linear regime (\(\Omega_p\!\ll\!\Omega_c\)); we keep a finite \(\Omega_p\) only to define \(\chi_{\rm exact}\).

\paragraph*{Step C — Grid and stability.}
We evaluate \(\chi_{\rm MF}\) on the same \(\Delta_1\)-grid used in the exact scan to allow pointwise comparison. The denominator is well-conditioned away from the two-photon pole by the finite \(\gamma_{2}/2\).

\subsection{Parameter sets}
\label{sec:numerics_params}

Unless stated otherwise, the EIT comparisons in the main text use:
\[
\begin{array}{l}
N=14,\quad
\Omega_p=0.1\Gamma,\quad
\Omega_c=0.5\Gamma,\quad
\Delta_2=0,\\
\Gamma_{31}=\Gamma_{32}=\Gamma,\quad
\gamma_{2}=\gamma_3=\gamma_\phi=10^{-4}\Gamma,\\
\Delta_1\in[-6,6],\quad \text{201 points}, \quad \text{Exact}\\
\Delta_1\in[-200,200],\quad \text{201 points}, \quad \text{MF}
\end{array}
\]
For SR transients (main text) we set \(\Omega_{p,c}=0\) and prepare an initially excited symmetric product state with a small ground-state admixture, then propagate the mean-field Hamiltonian dynamics as detailed there.

\subsection{Agreement metrics}
\label{sec:numerics_metrics}

To quantify the agreement between exact and MF susceptibilities we report, over the scan grid \(\{\Delta_{1,k}\}\),
\[
\varepsilon_{2}\equiv
\frac{\big\| f_{\rm exact}-f_{\rm MF}\big\|_{2}}{\big\| f_{\rm exact}\big\|_{2}},
\qquad
\varepsilon_{\infty}\equiv
\frac{\big\| f_{\rm exact}-f_{\rm MF}\big\|_{\infty}}{\big\| f_{\rm exact}\big\|_{\infty}},
\]
with \(f \in \{\mathrm{Im}\,\chi,\,\mathrm{Re}\,\chi\}\).
In addition, we extract physically meaningful single-point and shape errors:

\begin{itemize}
\item \textbf{On-resonance absorption:}
\(\varepsilon_{0}^{(\rm Im)}=\big|\mathrm{Im}\,\chi_{\rm MF}(0)-\mathrm{Im}\,\chi_{\rm exact}(0)\big|\).
\item \textbf{Line-center slope (dispersion):}
\(\varepsilon_{0}^{(\rm slope)}=\big|(\partial_{\Delta_1}\mathrm{Re}\,\chi)_{\rm MF}(0)-(\partial_{\Delta_1}\mathrm{Re}\,\chi)_{\rm exact}(0)\big|\),
evaluated by symmetric differences on the shared grid.
\item \textbf{EIT width and contrast:}
we define the EIT width \(\Delta_{\rm EIT}\) as the full width at half-\emph{dip} of \(\mathrm{Im}\,\chi(\Delta_1)\) around \(\Delta_1=0\)
and the contrast \(C_{\rm EIT}=1-\mathrm{Im}\,\chi(0)/\max_{\Delta_1\in\mathcal{W}} \mathrm{Im}\,\chi(\Delta_1)\),
where \(\mathcal{W}\) excludes a small interval around \(\Delta_1=0\).
Errors are \(\varepsilon_{\rm width}=|\Delta_{\rm EIT}^{\rm MF}-\Delta_{\rm EIT}^{\rm exact}|/\Delta_{\rm EIT}^{\rm exact}\) and
\(\varepsilon_{\rm contr}=|C_{\rm EIT}^{\rm MF}-C_{\rm EIT}^{\rm exact}|\).
\end{itemize}

\noindent
For SR transients (main text), we compare time series \(I(t)\), \(I_{13}(t)\), \(I_{23}(t)\) via the same norms.

\subsection{Reproducibility}
\label{sec:repro}

The numerical codes used to generate all figures and data in the main text and Supplementary Material are available from the authors upon reasonable request. This includes both the exact collective solver (QuTiP implementation) and the mean-field susceptibility scripts described in Section Numerical Methods. 
All simulations can be reproduced with standard Python~3.10+, NumPy, SciPy, Matplotlib, and QuTiP~4.7+.

\bibliographystyle{apsrev4-1}

\bibliography{References}

\end{document}